\documentclass[showpacs]{revtex4}
\usepackage{amsmath}
\usepackage{graphicx,rotating}

\begin{document}

\title{A Brownian Energy Depot Model of the Basilar
Membrane Oscillation with a Braking Mechanism}

\author{Yong Zhang}\email{xyzhang@phya.yonsei.ac.kr}
\author{Chul Koo Kim}\email{ckkim@yonsei.ac.kr}
\address{Institute of Physics and Applied Physics, Yonsei University,
Seoul 120-749, Korea,}
\author{Kong-Ju-Bock Lee}\email{kjblee@ewha.ac.kr}
\address{Department of Physics, Ewha Womans
University, Seoul 120-750, Korea}
\address{School of Physics, Korea Institute
for Advanced Study, Seoul 130-722, Korea}
\author{Youngah Park}\email{youngah@mju.ac.kr}
\address{Department of Physics, Myongji University, Yongin 449-728,
Korea}\address{School of Physics, Korea Institute for Advanced
Study, Seoul 130-722, Korea}


\begin{abstract}
High auditory sensitivity, sharp frequency selectivity, and
otoacoustic emissions are signatures of active amplification of the
cochlea. The human ear can also detect very large amplitude sound
without being damaged as long as the exposed time is not too long.
The outer hair cells are believed as the best candidate for the
active force generator of the mammalian cochlea. In this paper, we
propose a new model for the basilar membrane oscillation which
successfully describes both the active and the protective mechanisms
by employing an energy depot concept and a critical velocity of the
basilar membrane. One of the main results is that thermal noise in
the absence of external stimulation can be amplified leading to the
spontaneous basilar membrane oscillation. The compressive response
of the basilar membrane at the characteristic frequency and the
dynamic response to the stimulation are consistent with the
experimental results as expected. Our model also shows the nonlinear
distortion of the response of the basilar membrane.
\end{abstract}

\maketitle

\section{Introduction}
Remarkable detection capabilities appearing in animal hearing
are essentially governed by both a passive mechanical and an
active biophysical procedures in the
cochlea~\cite{Dallos,PhysiolRev_v81_1305,HC,JBP_v33_195}. The
active amplification of the living cochlea was conjectured by Gold
in 1948~\cite{PRSLB_v135_492} and is now qualitatively
studied widely through the Hopf equation in mathematical
models~\cite{PNAS_v97_3183,PRL_v90_158101,
PRL_v90_058101,PNAS_v101_12195,PRL_v93_268103}.
In order to give a better understanding to the experimental
observations, a more accurate model is required. Furthermore
the physical origin of the active oscillation remains
still unclear. In mammals, the outer hair cells (OHCs) are known
to be force generators for auditory sensitivity and frequency
selectivity. Proposed mechanisms of the force generation are
contraction of the OHC itself
~\cite{ARBE_v3_169,Nature_v419_300,CON_v13_459} and an
active motion of the hair bundle~\cite{PNAS_v98_14386,Neuron_v48_403}.

Energy depot model can describe the active
phenomena~\cite{Schweitzer,Zhang_NJP}, because the energy
supplied by the depot can induce a negative dissipation. A similar
mechanism for a negative stiffness in the bullfrog's hair bundle
has been reported~\cite{PNAS_v102_16996}. It is reasonable to consider the
OHC as an energy depot since the OHCs play an important role in the active
amplification of the cochlea even though we do not know the exact mechanism
of the energy supply. In nature all living things have some mechanisms
of protection for survival. We expect that there might be a similar
mechanism to
protect cochlea from damage by any external factors such as loud
stimulation. Hence, we propose a braking mechanism in the cochlea to
prevent the basilar membrane (BM) from an excessive oscillation which could
damage the cochlea. For this purpose, a critical velocity of the OHC
oscillation will be introduced in our model.

In the following
sections, we describe our model and reproduce the essential
phenomena of the active cochlea observed in recent experiments
such as the compressive nonlinearity, the dynamics of the BM
response, nonlinear distortion
and the spontaneous BM oscillation (SBMO). It will also
be pointed ut that our model naturally includes the Hopf
bifurcation model~\cite{PNAS_v97_3183,PNAS_v101_12195} in the case of weak
stimulation. However, in the regime of strong stimulation, the present
model is essentially different from the Hopf bifurcation model.

\begin{figure}[htbp]
\begin{center}
\includegraphics*[width=1.0\columnwidth]{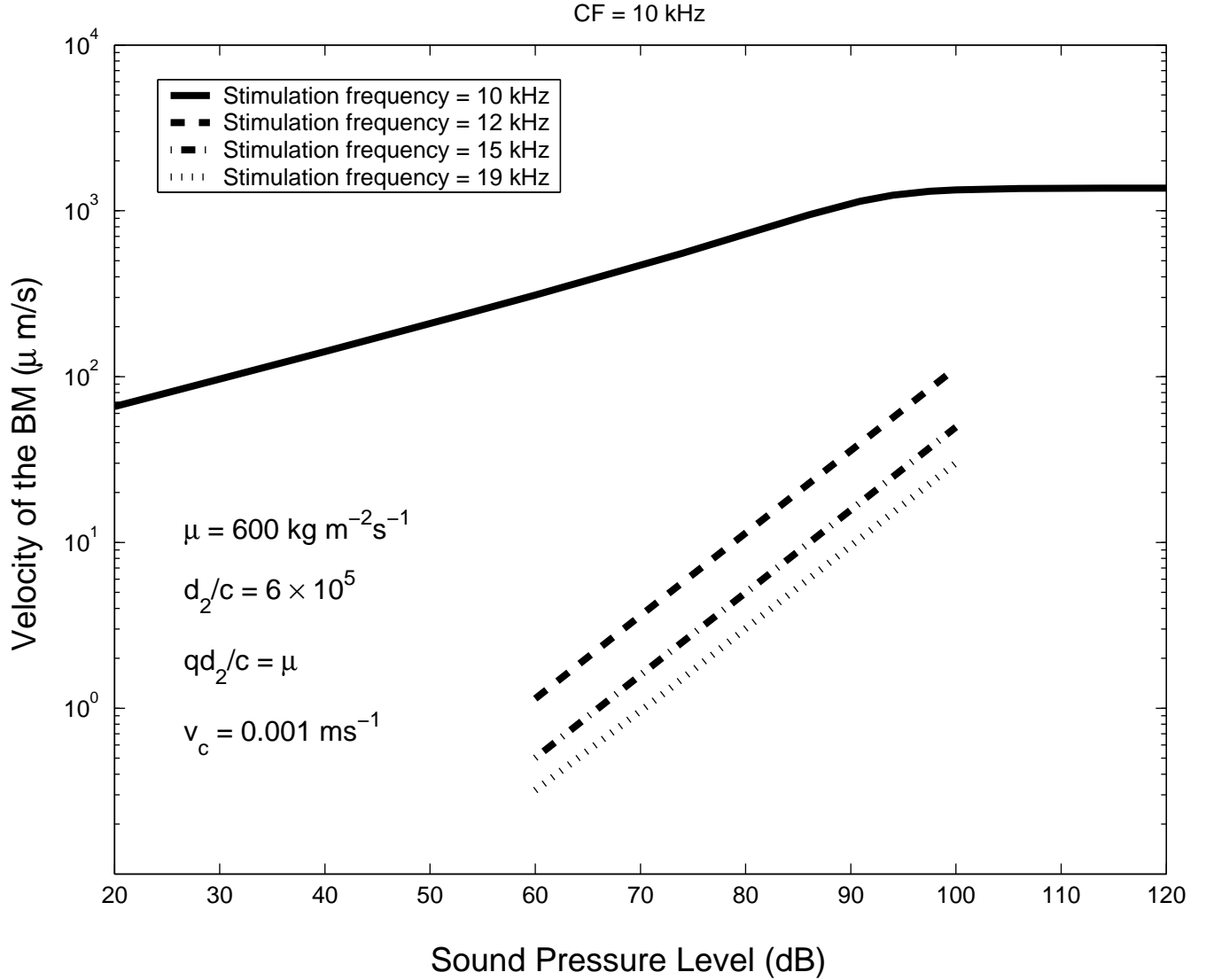}
\end{center}
\caption{The compressive response of the BM. The CF is $10$kHz. It
shows that the response of the BM is non-linear and compressive at
the CF. (Note that the slope of the solid line is less than one.)
Otherwise, the response of the BM is a linear function of the
stimulation yielding that the slopes are $1$. These results agree
with the experimental observations~\cite{JASA_v101_2151}.}
\label{compressive}
\end{figure}

\section{Energy Depot Model with Braking Mechanism}
It is well known that a change of transmembrane potential in
the OHC induces a somatic vibration. This implies that a part of the
electric energy is converted into the mechanical energy. In this sense, we
believe that the OHC functions as an energy depot for the BM. The OHC as
the energy depot stores energy supplied and converts it into kinetic energy
overcoming dissipation. This energetics can be described by the following
energy balance equation~\cite{Schweitzer}.
\begin{eqnarray}
\frac{dE(t)}{dt} = q - cE(t) - d(v)E(t),  \label{depot}
\end{eqnarray}
where $E(t)$ is the energy density of the energy depot, $q$ the
rate of the energy pumping into the energy depot per unit area,
$c$ the rate of the internal energy dissipation, $d(v)$ the rate
of the energy converting into the kinetic energy of the BM, and
$v$ the velocity of the BM. Considering that constant contribution
of $d(v)$ can be incooperated into $c$ and that conversion rate of the
energy into the kinetic energy of the BM will not be sensitive to
the sign of $v$, we assume that $d(v)$ is an even function of $v$
without a constant term. The energy depot model in which $d(v)=d_2
v^2$ has been extensively discussed~\cite{Schweitzer}. In this
paper, we include the next higher contribution to $d(v)$ and
introduce a critical velocity of the BM to describe a braking
mechanism~\cite{Zhang_NJP}.
Hence the conversion rate $d(v)$ can be written as
\begin{eqnarray}
d(v) = d_2 v^2 \left(1 - \frac{v^2}{v_c^2}\right), \label{ansaz}
\end{eqnarray}
where $v_c$ is the critical velocity. For a finite $v_c$ the
oscillation of the BM can be braked by the negative
conversion rate into the kinetic motion of the BM. Note that $v_c
\rightarrow \infty$ corresponds to a system without the braking
mechanism. The critical velocity will be roughly estimated from
the contraction of the OHC later for our numerical calculation.

\begin{figure}[htbp]
\begin{center}
\includegraphics*[width=1.0\columnwidth]{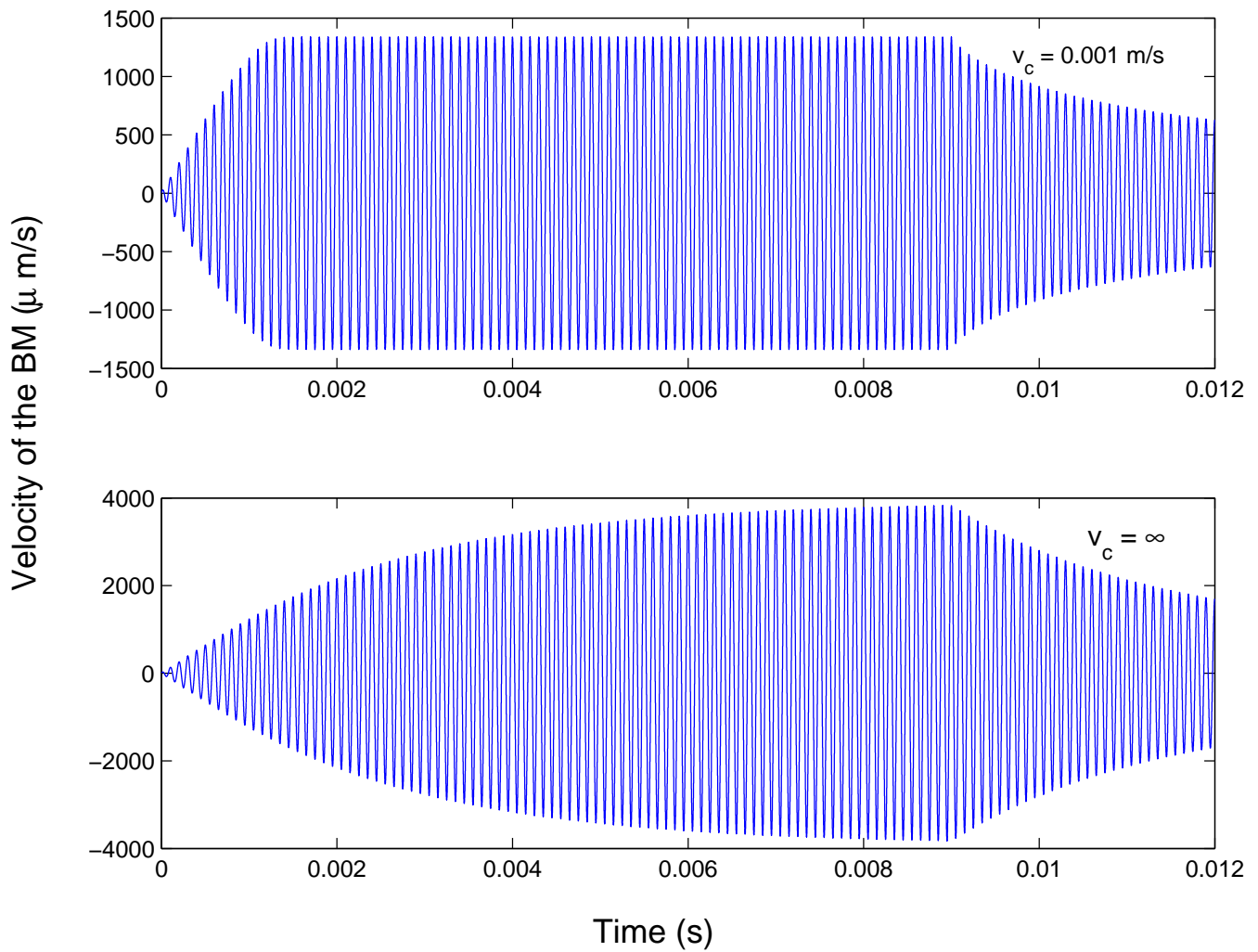}
\end{center}
\caption{The dynamic response of the BM when the CF is $10$kHz and
the stimulation is $100$dB. The onset time is much shorter than the
offset time which is not shown in full scale. This result is
qualitatively consistent with the experimental
observations~\cite{JASA_v101_2151}. $\mu = 600$kgm$^{-2}$s$^{-1}$,
$d_2/c = 6\times 10^5$m$^{-2}$s$^2$, and $qd_2/c = \mu$. The
stimulation is unloaded at $t = 9$ms. The lower panel shows the
response represents the case without the braking mechanism, showing
a continuous increase as long as the stimulation is on. }
\label{dyn}
\end{figure}

\begin{figure}[htbp]
\begin{center}
\includegraphics*[width=1.0\columnwidth]{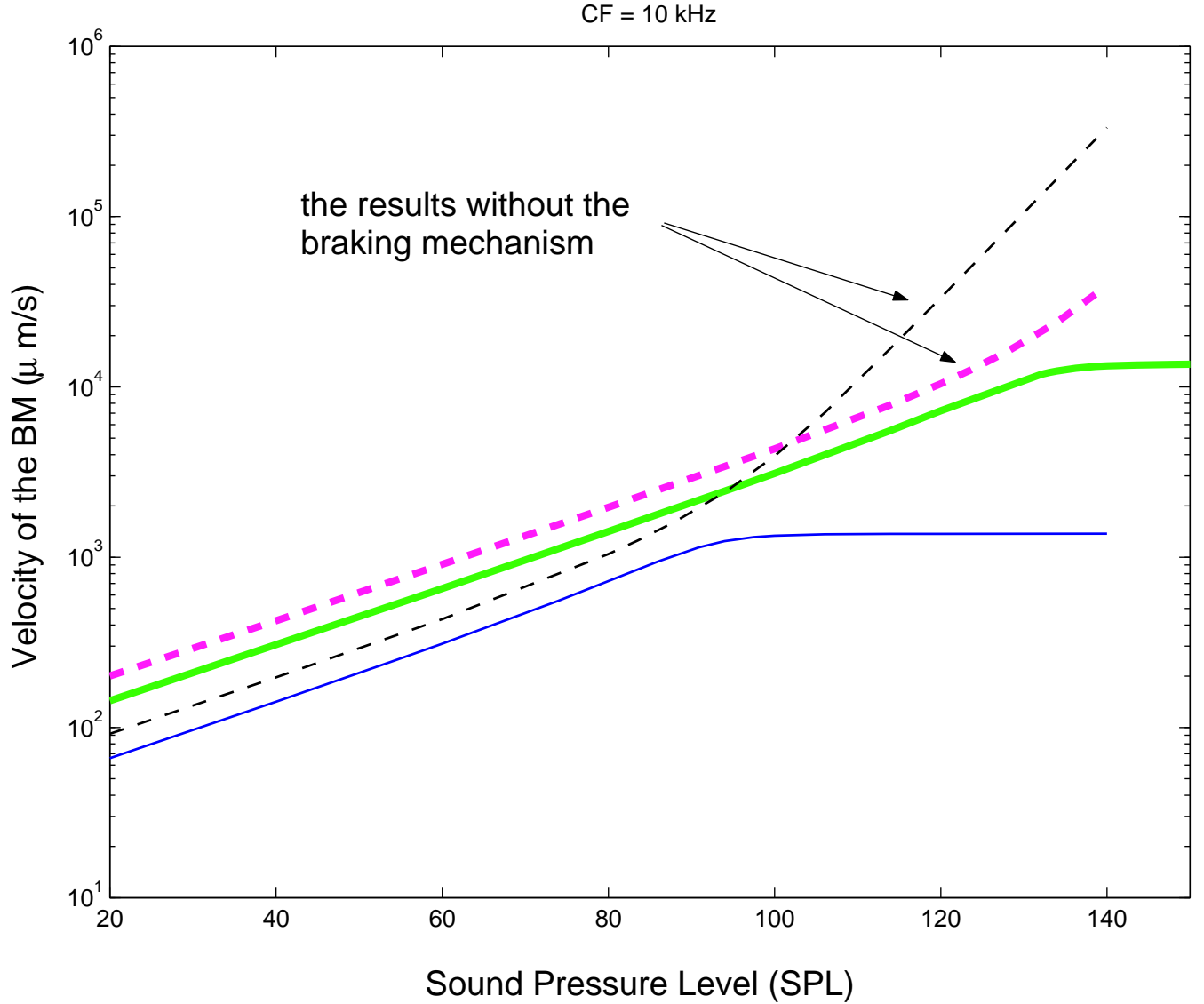}
\end{center}
\caption{(color online) The dependence of the response of the BM on
four different sets of parameters. {$\mu=600$kgm$^{-2}$s$^{-1}$,
$d_2/c = 6\times 10^5$m$^{-2}$s$^2$, $v_c = 0.001$m/s} (solid thin
line); {$\mu = 6,000$kgm$^{-2}$s$^{-1}$, $d_2/c =
6,000$m$^{-2}$s$^2$, $v_c = 0.01$m/s} (solid thick line); $\mu =
600$kgm$^{-2}$s$^{-1}$, $d_2/c = 6\times 10^5$m$^{-2}$s$^2$, $v_c =
\infty$ (dashed thin line), and {$\mu = 6,000 $kgm$^{-2}$s$^{-1}$,
$d_2/c = 6,000$m$^{-2}$s$^2$, $v_c =\infty$} (dashed thick line).
The velocity of the BM is (not) saturated when the stimulation is
sufficiently large with (without) the braking mechanism.}
\label{comp}
\end{figure}

Now it is worthwhile to compare our newly proposed model with the
one previously introduced in Ref.~\cite{PRL_v93_268103}. Whereas
we treat the energy density ($E(t)$) of the energy depot(OHC), in
Ref.~\cite{PRL_v93_268103} the energy density ($e(t)$) of the
propagating hydrodynamic wave is considered. The model in
Ref.~\cite{PRL_v93_268103} describes the coupling of the active
elements to the propagating wave, while the present model
describes the coupling between the energy depot(OHC) and the BM
oscillation. The emphasis in this paper is on showing that the
active and compressive response arises naturally from the
generalized energy depot concept.

\section{Results and Discussion}
An adiabatic approximation in which the adaptation of the energy depot is
very fast~\cite{Schweitzer} yields that
\begin{eqnarray}
E(t) = \frac{q}{c + d_2\left(1-v^2/v_c^2\right) v^2}.
\end{eqnarray}
It should bear in mind that $E(t)$ is not constant. The
oscillation of the BM is then governed by
\begin{eqnarray}
\rho\dot{v} + \mu v + \kappa R = F_{passive} + F_{active},
\label{full}
\end{eqnarray}
where $\rho$ is the density, $\mu$ the physical damping
coefficient, $\kappa$ the stiffness, and $R$
the displacement of the BM. $F_{passive}$ includes the
contributions of a sound wave (external stimulation) and a noise.
The noise is assumed to be a Gaussian white noise. $F_{active}$ is
the active force density acting on the BM by the OHC,
\begin{eqnarray}
F_{active} = \dfrac{d(v)E(t)}{v} =
\dfrac{q\dfrac{d_2}{c}\left(1-\dfrac{v^2}{v_c^2}\right)v}{1 +
\dfrac{d_2}{c}\left(1-\dfrac{v^2}{v_c^2}\right)v^2}. \label{active}
\end{eqnarray}
Note that the Langevin equation, Eq.~(\ref{full}), describes the
passive and the active response in a unified and natural way. It
should be noted that the active force, Eq.~(\ref{active}), can be
either positive or negative depending on the value of $v$. When
$v^2$ is smaller than $v_c^2$ (weak stimulation),
a positive active force is provided on
the BM (active mechanism). On the other hand, when $v^2$ is larger
than $v_c^2$ (strong stimulation),
a negative active force, hence an extra drag force,
is exerted on the BM (braking mechanism).

The mass density of the BM is known to be about
$0.77$kg/m$^2$~\cite{JASA_v112_576} and the reported damping
coefficient, $\mu$, of the BM has a large variation from $600$ to
$63,000$kg m$^{-2}$s$^{-1}$~\cite{JASA_v112_576, MMS_v4_664}
depending on the models. For a numerical calculation, we set the
damping in the range of $600 \sim 6,000$kg m$^{-2}$s$^{-1}$.
It has been reported that the Guinea pig OHCs are able
to change its length up to $5\%$ when the transmembrane potential
is varied. This corresponds to a displacement of $1\mu$m for the
OHCs of $20\mu$m at the high-frequency end~\cite{C7}. Hence,
in the regime of high frequency of $10$kHz, the adjustable
parameter $v_c$ is roughly estimated about $0.01$m/s.
By considering the OHC as a capacitor and using the observed
membrane potential, membrane capacitance~\cite{JNS_v11_3096},
and the scale of the OHC~\cite{C3}, we estimate the value of
$q$ to be about $0.01$kg s$^{-3}$. The response
of the BM to the sound wave can be then obtained straightforwardly
using Eqs.~(\ref{full}) and~(\ref{active}). In the following
calculations, we set $q d_2/c = \mu$ and the characteristic
frequency (CF), $\omega_c = \sqrt{\kappa/\rho}$. It will be shown later
that these relations arise from the conditions of the Hopf
bifurcation. In the present calculations, the noise width of the
distribution is chosen to be zero dB.

The compressive and dynamical responses of the BM are obtained as
shown in Fig.~\ref{compressive} and Fig.~\ref{dyn}, respectively.
Fig.~\ref{compressive} demonstrates that the response of the BM is
non-linear and compressive at the CF, while
the response is linear at other frequencies. This compressive
response of the BM agrees well with the
observations~\cite{PhysiolRev_v81_1305, JASA_v101_2151} as
expected. Fig.~\ref{dyn} shows the time course of the BM response,
reproducing the experimental observation that the
onset time of the response is shorter than the offset time,
although the onset time is somewhat larger than the experimental
values~\cite{JASA_v101_2151}. It is important to notice that the
velocity of the BM increases continuously as long as the
stimulation is being turned on if the braking mechanism is not included
$(v_c =\infty)$. Hence, the braking mechanism introduced in our
model is essential to explain the experimental observation at a
sufficiently strong stimulation. A dependence of the BM velocity
on $v_c$ is plotted in Fig.~\ref{comp}. The results clearly show
that lower critical velocity produces smaller response at the same
stimulation, thus producing a supercompressive behavior as observed in
experiment ~\cite{JASA_v101_2151}. This is the effect of the
braking mechanism, which is significantly different from the Hopf bifurcation
model. Note that the saturation of the BM response disappears if the
braking mechanism is not considered.

\begin{figure}[htbp]
\begin{center}
\includegraphics*[width=1.0\columnwidth]{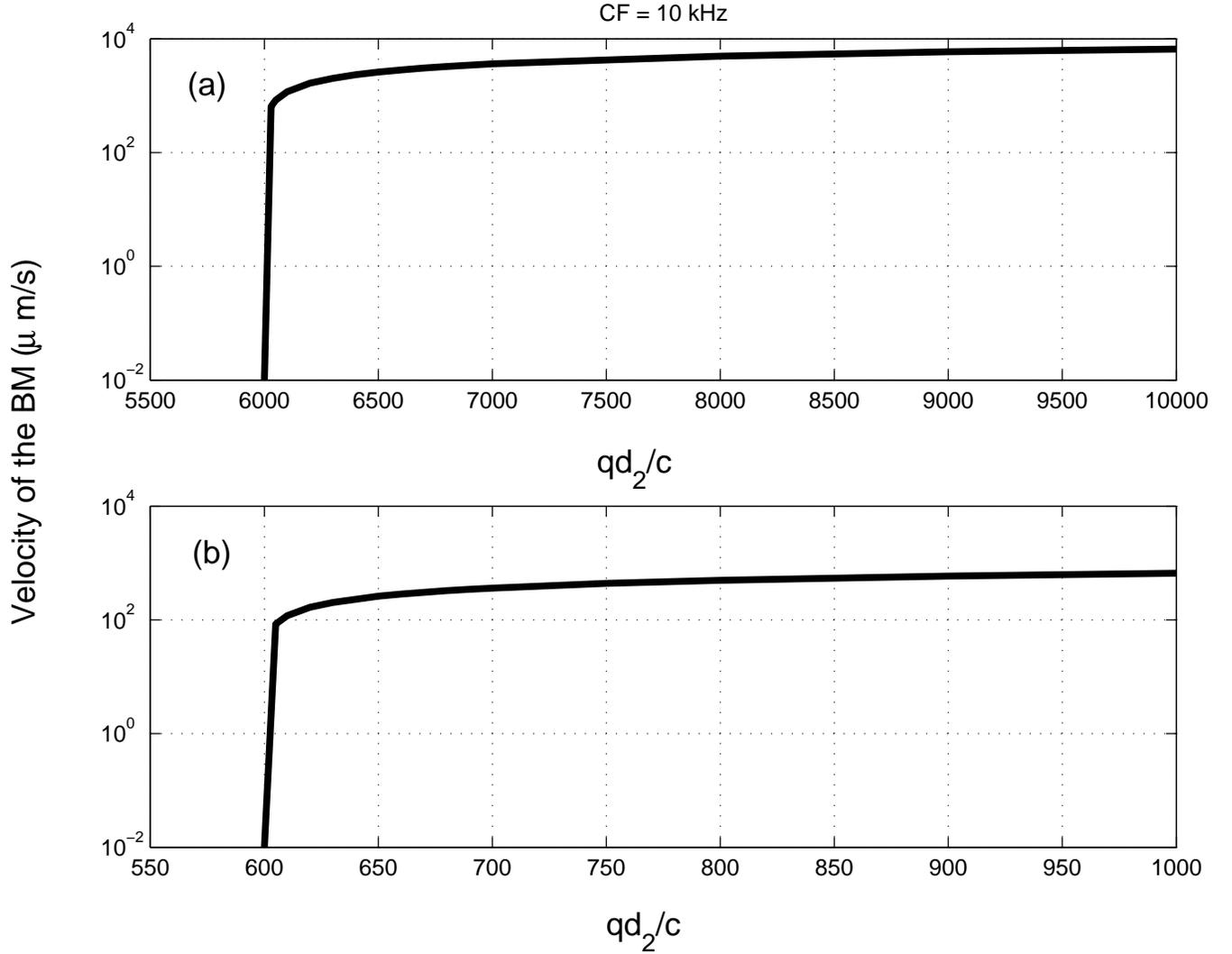}
\end{center}
\caption{The amplification of the thermal noise in the absence of
external stimulation. When $qd_2/c \leq \mu$, the Brownian
fluctuation is suppressed, while in the region of $qd_2/c
> \mu$, the Brownian fluctuation is amplified.  For the plots,
the intensity of the thermal noise is $0$Pa, and the strength of the
fluctuation is $0$dB. (a) $\mu=6,000$kgm$^{-2}$s$^{-1}$, $v_c =
0.01$m/s. (b) $\mu=600$kgm$^{-2}$s$^{-1}$, $v_c = 0.01$m/s.
}\label{vc}
\end{figure}

\begin{figure}[htbp]
\begin{center}
\includegraphics*[width=1.0\columnwidth]{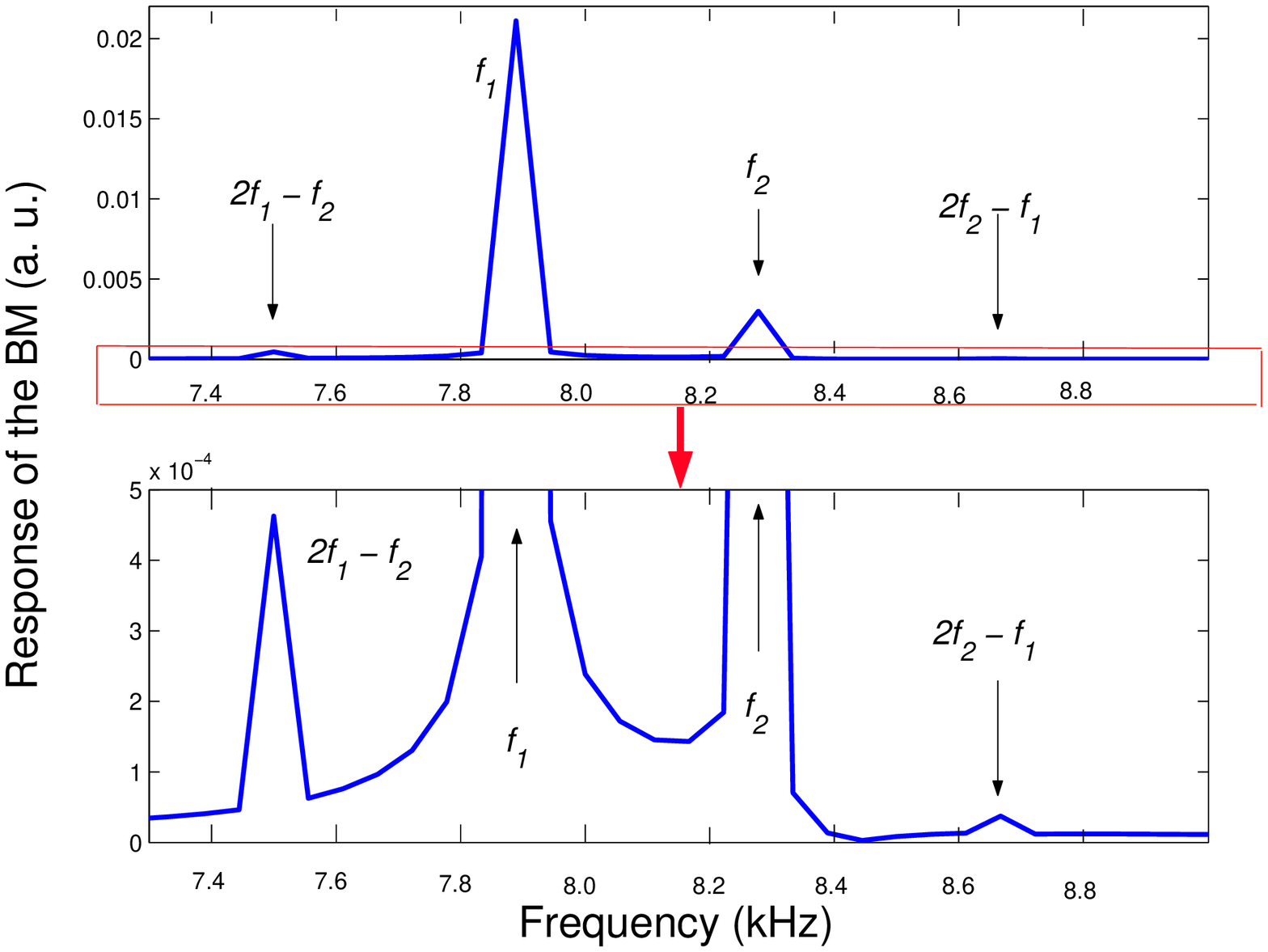}
\end{center}
\caption{Nonlinear distortion of the response of the BM. $f_1 =
7.89$kHz, $f_2 = 8.28$kHz, both strengths of the stimulation are
$100$dB, the response is measured at the place where the $CF =
7.89$kHz, $v_c^2 = 10^{-5}$m$^2$/s$^2$ and $\mu =
600$kgm$^{-2}$s$^{-1}$.} \label{dist}
\end{figure}

The gain of the oscillator is defined as the ratio of the
displacement of the BM to the stimulation,
\begin{eqnarray}
G = \frac{R}{F_{passive}}. \label{gain}
\end{eqnarray}
Using Eq.~(\ref{full}), we obtain the gain, which is a scaling
function of the stimulation when the stimulation is not too large,
$G \sim F_{passvie}^{-0.65}$. This result agrees well with the
experiment~\cite{PhysiolRev_v81_1305} and the result of the Hopf
bifurcation model~\cite{PNAS_v97_3183}, $G \sim
F_{passvie}^{-2/3}$. Indeed our model includes the Hopf model
naturally. To show this, we expand the active force,
Eq.~(\ref{active}), up to the lowest nonlinear term when $d_2(1 -
v^2/v_c^2)v^2/c$ is small. Eq.~(\ref{full}) is then approximately
written as
\begin{eqnarray}
F_{passive} &=& \rho \dot{v} + \left(\mu - q\frac{d_2}{c}\right)
v+\kappa R \nonumber \\
&+& q\frac{d_2}{c} \left(\frac{1}{v_c^2} + \frac{d_2}{c}\right) v^3.
\label{hp}
\end{eqnarray}
In the Fourier spectra space, this directly corresponds to the
Hopf equation~\cite{PNAS_v97_3183} with the following bifurcation
conditions,
\begin{eqnarray}
\omega_c &=& \sqrt{\frac{\kappa}{\rho}}, \label{cond1} \\
\mu &=& q\frac{d_2}{c}. \label{cond2}
\end{eqnarray}
Hence our energy depot model with a braking mechanism reduces to the
Hopf bifurcation model in the limit of weak stimulation.

One of the most important results of our model is an amplification
of the thermal noise itself in the absence of an external
stimulation. At the bifurcation point or when $qd_2/c < \mu$, the
thermal noise is well suppressed. However, when $qd_2/c > \mu$, so
that too much of the internal energy is converted into the kinetic
energy, the thermal noise can be largely amplified as shown in
Fig.~\ref{vc}. Here, we set the mean average of the noise at $0$Pa
and the width at $0$dB. Since it can be generally assumed that the
oscillators are distributed in the vicinity of the bifurcation
point, the noise may be amplified incoherently in the absence of
the input signal. However, when a weak input signal with the same
CF is introduced, it may phase-lock to the
already existing amplified noise, thus enhancing the selectivity
as discussed in Ref.~\cite{PNAS_v98_14380}. Because it is
believed that the SBMO is crucial for understanding the
spontaneous otoacoustic emission~\cite{JARO_v5_337}, our model
may provide a clue to explore this interesting and important
phenomenon. Fig.~\ref{dist} shows the nonlinear distortion of
the response of the BM at the position where the CF is $7.89$kHz. The
stimulation contains two tunes, $f_1 = 7.89$kHz and $f_2 = 8.28$kHz. The
strengths both are $100$dB. Our model shows clearly the distortions at
$2f_1 - f_2$, $2f_2 - f_1$ in the response of the BM.

\section{Summary}
In this paper, we propose a theoretical scheme for the active and
passive response of the BM based on a concept of the energy depot
model with a braking mechanism. According to the experimental
observations, the OHC is assumed to play a role
of the energy depot by supplying an active force to the BM and
also by reabsorbing an excessive kinetic energy like the regenerative-brake
system in automobiles and electric vehicles~\cite{ACC_v4_3129} .
It is shown that a single equation of
motion, Eq~(\ref{full}), can produce all the essential passive and
active behaviors. Moreover, our model shows that the Brownian
noise can be sufficiently amplified under a certain condition,
thus leading to the SBMO. Although we applied our model only to
the mammalian cochlea, it is equally possible to be applied to the
nonmammalian vertebrates~\cite{Nature_v433_880} because the only
required physical characteristics are a supply of the active force
and a self-adaptation. In this sense, we believe that the concept
of the energy depot model with a braking mechanism is quite
universal and can be applied to any underdamped biological
systems.

\section{Acknoledgement} This work is supported by the Korea Science
and Engineering Foundation (KOSEF) (R01-2006-000-10083-0).




\end{document}